\documentclass[draftclsnofoot, 12pt, onecolumn]{IEEEtran}

\usepackage[english]{babel}
\usepackage{amsmath,amssymb}
\usepackage{times}
\usepackage{relsize}
\usepackage{graphicx}
\usepackage{url}
\usepackage{booktabs}
\usepackage{multirow}
\usepackage{mparhack}
\usepackage{subfigure}
\usepackage{authblk}
\usepackage{bm}
\usepackage{algorithm}
\usepackage{algpseudocode}
\usepackage{epsfig}
\usepackage{epstopdf}
\usepackage{array}
\usepackage{cite}
\usepackage{stfloats}
\usepackage{balance}

\usepackage{amsmath}
\newcounter{subeqn} \renewcommand{\thesubeqn}{\theequation\alph{subeqn}}%
\makeatletter
\@addtoreset{subeqn}{equation}
\makeatother
\newcommand{\subeqn}{%
	\refstepcounter{subeqn}
	\tag{\thesubeqn}
}

\usepackage{epsfig}
\usepackage{xcolor}
\usepackage{tikz}

\newtheorem{theorem}{Theorem}

\newtheorem{proposition}[theorem]{Proposition}

\newtheorem{remark}{Remark}

\def\CS{{\mathcal S}}

\def\CA{{\mathcal A}}

\def\CA{{\mathcal A}}
\def\CAb{{\boldsymbol{\mathcal A}}}

\newcommand{\bs}{\boldsymbol}

\title{
Machine Learning-Enabled Joint Antenna Selection and Precoding Design: From Offline Complexity to Online Performance}

	\author{\IEEEauthorblockN{Thang X. Vu, \IEEEmembership{Member, IEEE}, Symeon Chatzinotas, \IEEEmembership{Senior Member, IEEE}, Van-Dinh Nguyen, \IEEEmembership{Member, IEEE}, Dinh Thai Hoang, \IEEEmembership{Member, IEEE}, Diep N. Nguyen, \IEEEmembership{Senior Member, IEEE}, Marco Di Renzo, \IEEEmembership{Fellow, IEEE}, and Bj\"orn Ottersten, \IEEEmembership{Fellow, IEEE}
		}\\
	\thanks{This work is supported by the European Research Council under project AGNOSTIC, grant 742648. Parts of this work were presented in the WiOPT 2019 \cite{VuWiOpt19}.}
	\thanks{T.~X.~Vu, S. Chatzinotas, V. D. Nguyen, and B. Ottersten are with the Interdisciplinary Centre for Security, Reliability and Trust (SnT) -- University of Luxembourg, L-1855 Luxembourg. Email: \{thang.vu, symeon.chatzinotas, dinh.nguyen, bjorn.ottersten\}@uni.lu.}
	\thanks{D. T. Hoang and D. N. Nguyen are with University of Technilogy Sydney, Sydney, Australia. Email: \{hoang.dinh, diep.nguyen\}@uts.edu.au.}
	\thanks{M. Di Renzo is with the Laboratoire des Signaux et Syst\`emes, CNRS, CentraleSupélec, Universit\'e Paris-Saclay, 91192 Gif-sur-Yvette, France. Email: marco.direnzo@centralesupelec.fr.}
}

\begin{document}

\date{}
\maketitle

\begin{abstract}
We investigate the performance of multi-user multiple-antenna downlink systems in which a base station (BS) serves multiple users via a shared wireless medium. In order to fully exploit the spatial diversity while minimizing the passive energy consumed by radio frequency (RF) components, the BS is equipped with $M$ RF chains and $N$ antennas, where $M < N$. Upon receiving pilot sequences to obtain the channel state information (CSI), the BS determines the best subset of $M$ antennas for serving the users. We propose a joint antenna selection and precoding design (JASPD) algorithm to maximize the system sum rate subject to a transmit power constraint and quality of service (QoS) requirements. The JASPD overcomes the non-convexity of the formulated problem via a doubly iterative algorithm, in which an inner loop successively optimizes the precoding vectors, followed by an outer loop that tries all valid antenna subsets. Although approaching the (near) global optimality, the JASPD suffers from a combinatorial complexity, which may limit its application in real-time network operations. To overcome this limitation, we propose a learning-based antenna selection and precoding design algorithm (L-ASPA), which employs a deep neural network (DNN) to establish underlaying relations between the key system parameters and the selected antennas. The proposed L-ASPD is robust against the number of users and their locations, BS's transmit power, as well as the small-scale channel fading. With a well-trained learning model, it is shown that the L-ASPD significantly outperforms baseline schemes based on the block diagonalization \cite{Chen07_AntenSelection} and a learning-assisted solution for broadcasting systems \cite{AS_multicast_Ibrahim2018} and achieves higher effective sum rate than that of the JASPA under limited processing time. In addition, we observed that the proposed L-ASPD can reduce the computation complexity by 95\% while retaining more than 95\% of the optimal performance. 
\end{abstract}


\section{Introduction}
Multiple-input multiple-output (MIMO) is an enabling technology to deal with the rapidly increasing demand for data-hungry applications in current and future mobile networks. By using a large number of antennas, an MIMO base station is able to send multiple information streams to multiple users simultaneously with negligible inter-user interference. The advantages of MIMO systems, under a proper beamforming design, comprise not only high spectral efficiency but also improved energy efficiency \cite{Larson2014}. When the number of antennas in MIMO systems becomes very large, antenna selection (AS) can be employed to improve the performance in terms of both hardware cost and technological aspects \cite{AS_Heath}. This is due to the fact that the radio frequency (RF) chains are usually much more expensive than antenna elements. More importantly, a proper AS strategy is capable of not only obtaining full spatial diversity but also considerably minimizing the RF chains' energy consumption, hence improving the system energy efficiency \cite{AS_Opt_Pei_2012}.
In general, AS is an NP-hard problem whose optimal solution is only guaranteed via exhaustive search, which tries all possible antenna combinations. The high complexity of AS may limit its potential in practice, especially in 5G services which usually have stringent latency and real-time decision making requirements \cite{5G}. 

Low-complexity solutions have become necessary to make AS practically feasible, especially for the BS of medium to large number of antennas. A block diagonalization-based algorithm is proposed in \cite{Chen07_AntenSelection} for multiuser MIMO systems, that selects the best antennas to either minimize the symbol error rate (SER) upper bound or maximize the minimum capacity. This method consecutively eliminates one antenna at a time that imposes the most energy in the corresponding orthogonal beamformers. The authors of \cite{AS_Multicast_norm12_TSP2013} propose a joint beamforming design and AS algorithm to minimize the multicasting transmit power. By using group sparsity-promoting $l_{1,2}$ norms instead of the $l_0$ norm, the selected antennas and beamformers can be obtained via an iterative algorithm. The application of $l_{1,2}$ norms is also employed in massive MIMO for minimizing the transmit power \cite{AS_MassiveMIMO_norm12_VTC2016} and in cell-free MIMO downlink setups for joint access point selection and power allocation \cite{VuICC20CF-MIMO}. In \cite{AS_broadcast_spawc18}, an AS algorithm based on mirror-prox successive convex approximation (SCA) is proposed for maximizing the minimum rate in multiple-input single-output (MISO) broadcasting systems. A similar SCA-based approach is proposed in \cite{Tervo2018,AS_EE_SPL16} for energy efficiency maximization.

Recently, the use of machine learning (ML) in communications systems has attracted much attention \cite{DeepL_PHY,Marco1, Marco2, leidlcaching,ML20_DLframeworkforMISO,ML19_FastBeamforming,ML20_FastBeamforming,ML19_LimitedfeedbackMIMO,ML20_BeamformingImperfectCSI_mmWave,ML20_CSInBeamforming,ML20_DRLBroadcast,ML20_DRLfor5G,ML_Dataset_mmWave}. The main advantage of ML-aided communications lies in the capability of establishing underlying relations between system parameters and the desired objective, hence being able to shift the computation burden in real-time processing to the offline training phase \cite{Sun_learntoopt,Lei_ICC18}. 
The authors of \cite{ML20_DLframeworkforMISO} propose a beamforming neural network (BNN) for minimizing the transmit power of multiuser MISO systems, which employs convolutional neural networks (CNN) and a supervised-learning method to predict the magnitude and direction of the beamforming vectors. This method is extended in \cite{ML19_FastBeamforming,ML20_FastBeamforming} for  unsupervised-learning to maximize the system weighted sum-rate. 
In \cite{ML19_LimitedfeedbackMIMO}, a deep learning-aided transmission strategy is proposed for single-user MIMO system with limited feed back, which is capable of addressing both pilot-aided training and channel code selection. 
The authors of \cite{ML20_BeamformingImperfectCSI_mmWave} develop a deep learning-based beamforming design to maximize the spectral efficiency of a single-user millimeter  wave (mmWave) MISO system, which achieves higher spectral efficiency than conventional hybrid beamforming designs.
The application of Q-learning is developed in \cite{ML20_CSInBeamforming} to overcome the combinatorial-complexity task of selecting the best channel impulse response in vehicle to infrastructure communications. 
A similar Q-learning based method is proposed in \cite{ML20_DRLfor5G} to solve the joint design of beamforming, power control, and interference coordination of cellular networks. 
In \cite{ML20_DRLBroadcast}, the authors develop a deep reinforcement learning framework which can autonomously optimize broadcast beams in MIMO broadcast systems based on users' measurements. 
A common data set for training mmWave MIMO networks is provided in \cite{ML_Dataset_mmWave} regarding various performance metrics.

Towards the learning-aided physical layer design, the application of ML to AS is a promising way to tackle the high-complexity of AS \cite{ML20_AS_HybridBeamformer,jingong,AS_multicast_Ibrahim2018,AS_wiretap_He2018}. A joint design for AS and hybrid beamformers for single-user mmWave MIMO is proposed in \cite{ML20_AS_HybridBeamformer} based on two serial CNNs, in which one CNN is used to predict the selected antennas and another CNN is used to estimate the hybrid beamformers. The authors of \cite{jingong} propose a multi-class classification approach to tackle the AS problem in single-user MIMO systems based on two classification methods, namely multiclass k-nearest neighbors and support vector machine (SVM). In \cite{AS_multicast_Ibrahim2018}, a neural network-based approach is proposed to reduce the computational complexity of AS for broadcasting. The neural network (NN) is employed to directly predict the selected antennas that maximize the minimum signal to noise ratio among the users. The authors of \cite{AS_wiretap_He2018} propose a learning-based transmit antenna selection to improve the security in the wiretap channel. Therein, two learning-based SVM and naive-Bayes schemes are considered. Although being able to improve the secrecy performance with a reduced feedback overhead, the setup analyzed in \cite{AS_wiretap_He2018} is limited to only a single antenna selection.


\subsection{Contributions}
In this paper, we investigate the performance of a multiuser MISO downlink system via a joint design of AS and precoding vectors to improve the system sum rate while guaranteeing the users' quality of service (QoS) requirements. Our contributions are as follows:
\begin{itemize}
	\item First, we develop a joint antenna selection and beamforming design (JASPD) framework to maximize the \emph{effective} system sum rate, which accounts for the time overhead spent on both channel estimation and computational processing, subject to users' QoS requirements and limited transmit power budget. The proposed JASPD works in an iterative manner, which first optimizes the beamforming vectors for a given antenna subset, and then selects the best antenna subset. 
	\item Second, to tackle the non-convexity in optimizing the beamforming vectors of JASPD, we propose two iterative optimization algorithms based on semidefinite relaxation (SDR) and SCA methods. The convergence of the proposed iterative algorithms to at least a local optimum is theoretically guaranteed.
	\item Third, we propose a learning-based antenna selection and precoding design (L-ASPD) algorithm to overcome the high computational complexity of AS, which employs a deep neural network (DNN) to capture and reveal the relationship between the system parameters and the selected antennas via an offline training process. More importantly, our leaning model is robust against not only the channel fading but also the number of users and their locations. Compared to existing works, which either study single-user MIMO systems \cite{ML20_AS_HybridBeamformer,jingong}, a single beamformer for broadcasting \cite{AS_multicast_Ibrahim2018} or a single antenna selection \cite{AS_wiretap_He2018}, we consider a more general multi-user system.
	\item Finally, extensive simulation results show that, under the same limited processing time, the proposed L-ASPD outperforms the JASPD and significantly outperforms existing AS schemes on both model-based \cite{Chen07_AntenSelection} and ML-aided \cite{AS_multicast_Ibrahim2018} designs. We observed that the L-ASPD can achieve more than 95\% of the optimal sum rate while reducing more than 95\% of the computational time.
\end{itemize}

The rest of the paper is organized as follows. Section II presents the system model and key parameters. 
Section III develops two iterative optimization algorithms used in the JASPD. Section IV introduces a ML-aided joint design to accelerate real-time processing. Section V demonstrates the effectiveness of the proposed algorithms via simulation results. Finally, Section IV concludes the paper.

\emph{Notations}: The superscript $(.)^T$, $(.)^H$ and $\mathrm{Tr}(.)$ stand for the transpose, Hermitian transpose, and trace operation, respectively.  $\binom{n}{k}$ represents the binomial coefficients. $|.|$ and $\|.\|$ denote the cardinality and the $l_2$-norm of a set, respectively.

\section{System Model}

\begin{figure}
	\centering
	\includegraphics{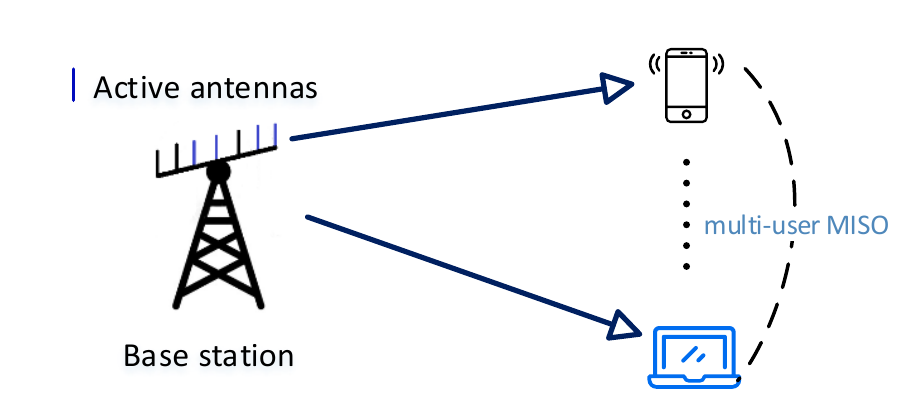} 
	\caption{Diagram of multiuser MISO system. A subset of antennas is selected for data transmission.}
	\label{fig:systemmodel}
\end{figure}

We consider a multiuser MISO downlink system operated in time division duplex (TDD) mode, in which a multi-antenna base station (BS) servers $K$ single-antenna users in the same frequency resource\footnote{In practice the whole bandwidth is divided into multiple sub-frequency bands. The proposed scheme is directly applied to each band.}, as depicted in Fig.~\ref{fig:systemmodel}. The BS is equipped with $M$ RF chains and $N$ antennas, where $N > M \ge K$. The motivation of having more antennas than the number of RF chains is that the BS can i) fully exploit spatial diversity gain and ii) minimize the static energy consumed by hardware components \cite{AS_Opt_Pei_2012}, e.g., RF chains and amplifiers.
The system operates in a quasi-static block fading channel in which the channel gains are constant within on block and independently change from one block to another. Before sending data to the users, the BS needs to acquire the channel state information (CSI) via pilot-aided channel estimation\footnote{The system is assumed to operate above certain SNR levels in which the CSI can be efficiently estimated.} in order to perform reprocessing, e.g., beamforming and power allocation. 

\begin{figure}
	\centering
	\includegraphics[width = 0.8\columnwidth]{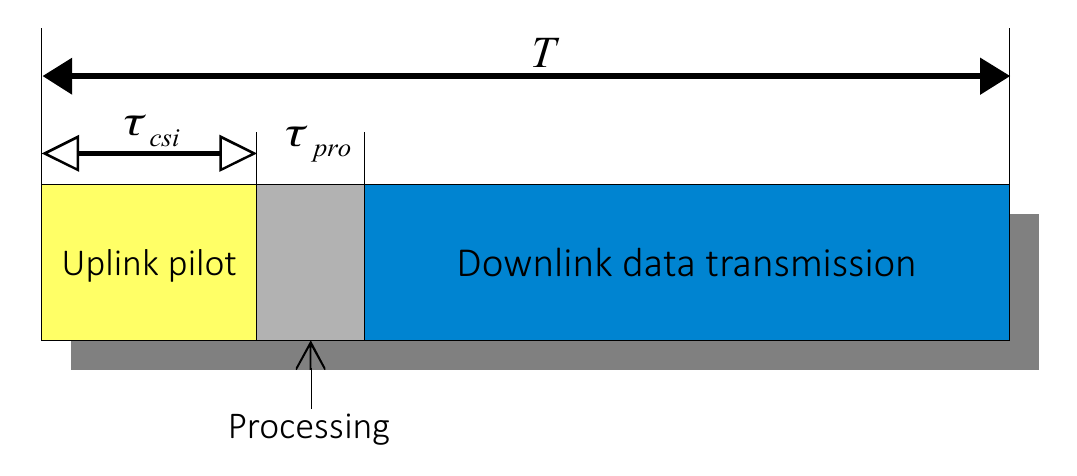}
	\caption{Block diagram of one transmission block.}
	\label{fig:tdd}
\end{figure}

Fig.~\ref{fig:tdd} illustrates the three phases in one transmission block. Let $T$ and $\tau_{csi}$ denote the block duration and channel estimation time, both expressed in terms of channel use (c.u.), respectively. The block duration is determined by the system coherence time. Assuming mutually orthogonal pilot sequences across the users, the channel estimation time is $\tau_{csi} = K(\lfloor N/M\rfloor + 1)$ c.u., where $\lfloor x\rfloor$ denotes the largest integer not exceeding $x$. Unlike most of previous works that ignore the processing time, we consider the general case in which the processing time takes place in $\tau_{pro}$ (c.u.). In practice, the value of $\tau_{pro}$ largely depends on beamforming techniques and the hardware capability.

Let $\bs{h}_k \in \mathbb{C}^{1\times N}$ denote the channel vector from the BS's antennas to user $k$, including the pathloss.
We assume that full CSIs are available at the BS. Because there are only $M < N$ RF chains, the BS has to determine an optimal subset of $M$ antennas for sending data to the users.  Let $\CA = \{a_1, a_2, \dots, a_M\}, a_m\in [N]\triangleq \{1, 2, \dots, N\}$, be a subset of $M$ antennas (out of $N$), and let $\CAb$ be the collection of all possible antenna subsets. By definition, we have $|\CA| = M$ and $|\CAb| = \binom{N}{M}$. 

Denote by $\bs{h}_{k,\CA} \in \mathbb{C}^{1\times M}$ the channel vector from active antennas in a subset $\CA$ to user $k$, i.e., $\bs{h}_{k,\CA} = \left[h_k[a_1], h_k[a_2], \dots, h_k[a_M]\right]$, where $a_m \in \CA$ and $h_k[n]$ is the $n$-th element of $\bs{h}_k$. Before serving the users, the BS first precodes the data to suppress inter-user interference. Let $\bs{w}_{k,\CA} \in \mathbb{C}^{M\times 1}$ be the precoding vector for user $k$ corresponding to the selected antenna subset $\CA$. The received signal at user $k$ is 
\begin{align}
	y_{k,\CA} = \bs{h}_{k,\CA} \bs{w}_{k,\CA} x_k + {\sum}_{i\neq k} \bs{h}_{k,\CA} \bs{w}_{i,\CA} x_i + n_k, \label{eq:y_k}
\end{align}
where $n_k$ is Gaussian noise with zero mean and variance $\sigma^2$. The first term in \eqref{eq:y_k} is the desired signal, and the second term is the inter-user interference.

By considering interference as noise, the \emph{effective achievable rate} of user $k$ is 
\begin{align}
	R_k(\CA) =& B\left(1 - \frac{\tau_{csi} + \tau_{pro}}{T} \right) \notag \\
	&\times \log_2\Big(1 + \frac{|\bs{h}_{k,\CA} \bs{w}_{k,\CA}|^2}{{\sum}_{i \neq k} |\bs{h}_{k,\CA} \bs{w}_{i,\CA}|^2 + \sigma^2} \Big), \forall k, \label{eq:R_k}
\end{align}
where $B$ is the shared channel bandwidth and $1 - \frac{\tau_{csi} + \tau_{pro}}{T}$ accounts for actual time for data transmission. The total transmit power\footnote{The energy consumed by hardware components is excluded since it is constant and does not affect the precoding design.} is ${\sum}_{k=1}^K\|\bs{w}_{k,\CA} \|^2$. 


\begin{remark}
	It is observed from \eqref{eq:R_k} that the effective data rate is determined not only by the precoding vectors $\bs{w}_{k,\CA}$ but also by the channel estimation and processing times. In particular, spending more time on either channel estimation or processing will degrade the effective transmission rate. 
\end{remark}
\section{Optimal Antenna Selection and Precoding Design}\label{sec:JASPD}
In this section, we develop a joint antenna selection and precoding design to maximize the system sum rate while satisfying the minimum QoS requirements and limited power budget. The joint optimization problem can be formulated as follows:
\begin{align}
\text{P0}:~&\underset{\CA\in \CAb, \{\bs{w}_{k,\CA}\}}{\mathtt{maximize}}
~~ {\sum}_{k=1}^K R_k(\CA) \label{OP:0} \\
\mathtt{s.t.} 
&~~ R_k(\CA) \ge \eta_k, \forall k, \notag \\
&~~ {\sum}_{k=1}^K \|\bs{w}_{k,\CA} \|^2\leq P_{tot}, \notag 
\end{align}
where $R_k(\CA)$ is given in \eqref{eq:R_k}, $P_{tot}$ is the total transmit power budget at BS, and $\eta_k$ is the QoS requirement for user $k$. In problem \eqref{OP:0}, the first constraint is to satisfy the minimum user QoS requirement and the second constraint states that the total transmit power should not exceed the power budget. We note that the problem formulation in \eqref{OP:0} can be directly extended to the weighted sum rate metric for given weight coefficients with the weights are used as parts of the training input.
 
In general, problem \eqref{OP:0} is a mixed binary non-linear problem where the binary variables of the activated antennas are strongly coupled with the continuous variables of the precoding vectors. Because the precoding vectors are designed for a given selected antenna subset, problem {P0} can be reformulated in an iterative form as follows:
\begin{align}
	\underset{\CA \in \CAb}{\mathtt{maximize}} ~~ \text{P1}(\CA), \label{OP:0 AS} 
\end{align}
where $\text{P1}(\CA)$ is the precoding design problem for the candidate antenna subset $\CA$, which is defined as follows
\begin{align}
&\mathrm{P1}(\CA): \underset{\{\bs{w}_{k,\CA}\}}{\mathtt{Max}}
~~ \bar{B}\sum_{k=1}^K \log_2\Big(1 + \frac{|\bs{h}_{k,\CA} \bs{w}_{k,\CA}|^2}{{\sum}_{i \neq k} |\bs{h}_{k,\CA} \bs{w}_{i,\CA}|^2 + \sigma^2} \Big) \label{OP:1}\\
&\mathtt{s.t.} 
~~ \bar{B}\log_2\Big(1 + \frac{|\bs{h}_{k,\CA} \bs{w}_{k,\CA}|^2}{{\sum}_{i \neq k} |\bs{h}_{k,\CA} \bs{w}_{i,\CA}|^2 + \sigma^2} \Big) \ge \eta_k,\forall k,\subeqn \label{eq:p1 c1}\\
&\qquad~~{\sum}_{k=1}^K \|\bs{w}_{k,\CA} \|^2\leq P_{tot}, \subeqn \label{eq:p1 c2}
\end{align}
where $\bar{B} \triangleq B(1 - \frac{\tau_{csi}+\tau_{pro}}{T})$ and we have used \eqref{eq:R_k} for $R_k(\CA)$.

If problem $\text{P1}(\CA)$ can be solved optimally, then the optimal solution of P0 can be obtained via an exhaustive search in \eqref{OP:0 AS}, which tries all possible antenna subsets. Unfortunately, solving problem $\text{P1}(\CA)$ is challenging due to the non-concavity of the objective function and the non-convexity of the first constraint. 

In the following, we propose two solutions based on SDR and SCA methods to tackle the non-convexity of the beamforming vectors design in Section~III-A. We then describe the proposed JASPD algorithm and analyze its complexity in Section~III-B.

\subsection{Near Optimal Beamforming Design for Selected Antennas}
In this subsection, we design the beamforming vectors to maximize the system sum rate for a selected antenna subset. In the following, we propose two methods to solve \eqref{OP:1}.

\subsubsection{Semidefinite Relaxation based Solution}
Semidefinite-based formulation is an efficient method to design the beamforming vectors of wireless systems, which converts quadratic terms into linear ones by lifting the original variable domain into a higher-dimensional space. We adopt the semidefinite method to deal with the signal-to-noise-plus-interference-ratio (SINR) term in both the objective function and the first constraint. Define a new set of variables $\bs{W}_{k} = \bs{w}_{k,\CA}\bs{w}^H_{k,\CA} \in \mathbb{C}^{M\times M}$, and denote $\bs{H}_{k} \triangleq \bs{h}^H_{k,\CA} \bs{h}_{k,\CA}$. It is straightforward to verify that $|\bs{h}_{k,\CA}\bs{w}_{l,\CA}|^2 = \bs{h}_{k,\CA}\bs{w}_{l,\CA} \bs{w}^H_{l,\CA} \bs{h}^H_{k,\CA} = \mathrm{Tr}(\bs{H}_{k}\bs{W}_{l})$ and $\|\bs{w}_{k,\CA}\|^2 = \mathrm{Tr}(\bs{W}_{k})$.

By introducing arbitrary positive variables $\{x_k\}_{k=1}^K$, we can reformulate problem \eqref{OP:1} as follows:
\begin{align}
&\underset{\bs{W}, \bs{x}}{\mathtt{maximize}}
~~ \frac{\bar{B}}{\log(2)}{\sum}_{k=1}^K x_k  \label{OP:SDR1} \\
&\mathtt{s.t.} 
~~ \log\Big(1 + \frac{\mathrm{Tr}(\bs{H}_k\bs{W}_k)}{{\sum}_{i\neq k} \mathrm{Tr}(\bs{H}_k\bs{W}_i) + \sigma^2}\Big) \ge  x_k, \forall k,  \subeqn \label{eq:OP SDR1 c1}\\
&\qquad~~  x_k \ge \frac{\eta_k\log(2)}{\bar{B}},\forall k, \subeqn \label{eq:OP SDR1 c2} \\
&\qquad~~ {\sum}_{k=1}^K\mathrm{Tr}(\bs{W}_k) \le P_{tot},\subeqn \label{eq:OP SDR1 c3}\\
&\qquad~~ \mathrm{rank}(\bs{W}_k) = 1, \forall k,\notag
\end{align}
where we use short-hand notations $\bs{W}$ and $\bs{x}$ for $(\bs{W}_1, \dots, \bs{W}_K)$ and $(x_1, \dots, x_K)$, respectively.

The equivalence between \eqref{OP:SDR1} and \eqref{OP:1} can be verified as the equality holds in \eqref{eq:OP SDR1 c1} at the optimum. It is observed that the objective is a linear function and constraints \eqref{eq:OP SDR1 c2} and \eqref{eq:OP SDR1 c3} are convex. Thus, the challenge in solving problem \eqref{OP:SDR1} lies in \eqref{eq:OP SDR1 c1} and the rank-one constraint. While the latter constraint can be efficiently coped with by using the relaxation method followed by randomization if needed \cite{Luo2010}, dealing with the former constraint is more struggling.

In the next step, we introduce slack variables $\{y_k\}_{k=1}^K$ and reformulate constraint \eqref{eq:OP SDR1 c1} as 
\begin{align}
    &\log\Big(\sigma^2 + {\sum}_{i=1}^K \mathrm{Tr}(\bs{H}_k\bs{W}_i)\Big) \ge x_k + y_k, \label{eq:app1}\\
    &\sigma^2 + {\sum}_{i\ne k} \mathrm{Tr}(\bs{H}_k\bs{W}_i) \le e^{y_k}. \label{eq:app2}
\end{align}
%
%
Because the function $\log()$ is concave, constraint \eqref{eq:app1} is convex. However, since the function $\exp(.)$ is convex, constraint \eqref{eq:app2} is unbounded. To overcome this difficulty, we employ the inner approximation method, which uses the first-order approximation of $e^{y_k}$ at the right hand side of \eqref{eq:app2}. As a result, the approximated problem of \eqref{OP:SDR1} can be formulated as follows:
\begin{align}
&\mathrm{P2}(\bs{y}_0):~ \underset{\bs{W},\bs{x}, \bs{y}}{\mathtt{maximize}}
~~ \frac{\bar{B}}{\log(2)}{\sum}_{k=1}^K x_k  \label{OP:SDR1 app} \\
&\mathtt{s.t.} 
~~ \eqref{eq:OP SDR1 c2}; \eqref{eq:OP SDR1 c3}; \eqref{eq:app1};~\mathrm{rank}(\bs{W}_k) = 1, \forall k, \notag \\
&~~  \sigma^2 + {\sum}_{i\neq k} \mathrm{Tr}(\bs{H}_k\bs{W}_i) \le 
e^{y_{0k}}(y_k - y_{0k} + 1), \forall k, \subeqn \label{eq:OP SDR1 app c1}
\end{align}
where $\bs{y} \triangleq \{y_{k}\}_{k=1}^K$ and $\bs{y}_0$ is any feasible value of $\bs{y}$ that satisfies constraint \eqref{eq:app2}. 

It is evident that, for a given $\bs{y}_0$, the objective and constraints of problem \eqref{OP:SDR1 app} are convex except for the rank one constraint. This suggests to solve \eqref{OP:SDR1 app} by the semi-definite relaxation (SDR) method \cite{Luo2010} which ignores the rank one constraint and can be solved in an efficient manner by standard solvers, e.g., CVX. Because $e^{y_0}(y - y_0 + 1) \leq e^{y}, \forall y_0$, the approximated problem \eqref{OP:SDR1 app} always gives a suboptimal solution of the original problem \eqref{OP:SDR1}.

\begin{algorithm}[t]
	\caption{\textsc{Iterative Algorithm to solve \eqref{OP:SDR1}}}\label{algo:SDR}
	\begin{algorithmic}[1]
		\State Initialize $\bs{y}_0$, $\epsilon$, $X_{\rm old}$ and $\mathtt{error}$.
		\State \textbf{while} $\mathtt{error} > \epsilon$ \textbf{do} 
		\State  \qquad Solve the SDR of \eqref{OP:SDR1 app} by dropping the rank-one constraint to obtain  $\{\bs{W}_{\star k}, x_{\star k}, y_{\star k}\}_{k=1}^K$
		\State \qquad Compute $\mathtt{error} = \frac{\bar{B}}{\log(2)}|\sum_{k=1}^Kx_{\star k} - X_{\rm old}|$
		\State \qquad Update $X_{\rm old} \gets \frac{\bar{B}}{\log(2)}\sum_{k=1}^Kx_{\star k}; y_{0k} \gets y_{\star k}, \forall k$
	\end{algorithmic} 
\end{algorithm}
It is worth noting that the optimal solution of problem \eqref{OP:SDR1 app} is largely determined by the parameters $\bs{y}_0$. Thus, it is crucial to select proper values $\bs{y}_0$ such that the solution of \eqref{OP:SDR1 app} is close to the optimal solution of \eqref{OP:SDR1}. As such, we propose an iterative optimization algorithm to improve the performance of problem \eqref{OP:SDR1 app}, shown in Algorithm~\ref{algo:SDR}. The premise behind the proposed algorithm is to better estimate $\bs{y}_0$ through iterations. 
\begin{proposition}[Convergence of Algorithm~\ref{algo:SDR}]\label{prop SDR}
	The sequence of the objective values generated by Algorithm 1 in solving  the SDR of problem P2($\bs{y}_0$) is non-decreasing.
\end{proposition}
The proof of Proposition~\ref{prop SDR} is shown in Appendix~\ref{app:prop SDR}. Although not guaranteeing the global optimum of problem \eqref{OP:SDR1}, Proposition~\ref{prop SDR} justifies the convergence to at least a local optimum of the proposed iterative algorithm\footnote{The study of the performance gap to the global optimum is left for future work.}.

\begin{remark}[Initialization of Algorithm~\ref{algo:SDR}]
	The execution of Algorithm~\ref{algo:SDR} requires initial values $y_{0k}, \forall k$. Therefore, it requires an efficient way to find these initial values before tackling problem \eqref{OP:SDR1 app}. To this end, we start by solving the feasibility problem below:
	 \begin{align}
	{\mathtt{Find}}
	 &~~ \bs{W} \label{OP:SDR1 ini} \\
	 \mathtt{s.t.} 
	 &~~ \frac{\mathrm{Tr}(\bs{H}_k\bs{W}_k)}{2^{\eta_k/\bar{B}}-1} \ge {\sum}_{i\neq k} \mathrm{Tr}(\bs{H}_k\bs{W}_i) + \sigma^2, \forall k, \notag\\
	 &~~  {\sum}_{k=1}^K\mathrm{Tr}(\bs{W}_k) \le P_{tot}, \notag
	 \end{align}
	 which is convex. Then the initial values are computed as $y_{0k} = \log({\sum}_{i\neq k} \mathrm{Tr}(\bs{H}_k\bs{W}^*_i) + \sigma^2 ), \forall k$, where $\bs{W}^*_k$ is the solution of \eqref{OP:SDR1 ini}.
\end{remark}
\begin{remark}[Randomization]
	The solution in \eqref{OP:SDR1 app} is based on the SDR which sometimes violates the rank-one constraint. In such cases, Gaussian randomization can be adopted. Details on Gaussian randomization process are available in \cite{Luo2010}. Our simulation results show that more than 99\% of the times Algorithm~\ref{algo:SDR} can output rank-one solutions.	
\end{remark}
\subsubsection{Reformulation based on Difference of Convex}
The SDR-based reformulation in the previous subsection leverages the original problem's non-convexity by working in a higher dimensional domain, which requires more memory. In this subsection, we solve \eqref{OP:1} based on difference-of-convex (DC) reformulation directly on the original variable domain. 

By introducing arbitrary positive variables $\bs{u} \triangleq \{u_k\}_{k=1}^K$, we can reformulate problem \eqref{OP:1} as follows:
\begin{align}
\underset{\bs{w}, \bs{u}}{\mathtt{Maximize}}
&~~ \bar{B}{\sum}_{k=1}^K \log_2( 1 + u_k)  \label{OP:dc1} \\
\mathtt{s.t.} 
&~~ \frac{|\bs{h}_{k,\CA} \bs{w}_{k,\CA}|^2}{{\sum}_{i \neq k} |\bs{h}_{k,\CA} \bs{w}_{i,\CA}|^2 + \sigma^2} \ge u_k, \forall k, \subeqn \label{eq:OP dc1 c1}\\
&~~  u_k \ge \bar{\eta}_k, \forall k, \subeqn \label{eq:OP dc1 c2} \\
&~~{\sum}_{k=1}^K \|\bs{w}_{k,\CA} \|^2\leq P_{tot}, \subeqn \label{eq:OP dc1 c3}
\end{align}
where $\bar{\eta}_k \triangleq 2^{\eta_k/\bar{B}}-1$ and $\bs{w}$ is a short-hand notation for $(\bs{w}_{1,\CA}, \dots, \bs{w}_{K,\CA})$. 
The equivalence between \eqref{OP:dc1} and \eqref{OP:1} can be verified since constraint \eqref{eq:OP dc1 c1} holds with equality at the optimum. 

As the denominator of the left-hand-side of \eqref{eq:OP dc1 c1} is positive, it can be rewritten as
\begin{align}
\frac{|\bs{h}_{k,\CA} \bs{w}_{k,\CA}|^2}{u_k} \ge {\sum}_{i \neq k} |\bs{h}_{k,\CA} \bs{w}_{i,\CA}|^2 + \sigma^2. \label{eq:OP2 c1 equiv}
\end{align}

An important observation from \eqref{eq:OP2 c1 equiv} is that $\frac{|\bs{h}_{k,\CA} \bs{w}_{k,\CA}|^2}{u_k}$ is a convex function of $\bs{w}_{k,
\CA}$ and $u_k$ (see Appendix~~\ref{app 1}). Therefore, \eqref{eq:OP2 c1 equiv} has a form of the DC representation, which suggests an efficient way to solve \eqref{eq:OP dc1 c1}. In particular, let $\hat{\bs{w}}_{k,\CA}, \hat{u}_k$ be any feasible solution of \eqref{OP:dc1}, we can approximate \eqref{eq:OP2 c1 equiv} by using the first order approximation of the left-hand-side of \eqref{eq:OP2 c1 equiv}, stated as
\begin{align}
	&\sum_{i \neq k} \bs{w}^H_{k,\CA}\bs{H}_k \bs{w}_{i,\CA} + \sigma^2 \le \frac{\bs{w}^H_{k,\CA}\left(\!\bs{H}_k\! +\! \bs{H}^T_k\!\right) \hat{\bs{w}}_{k,\CA}}{\hat{u}_k} \notag\\
	&- u_k \frac{\hat{\bs{w}}^H_{k,\CA}\bs{H}_k \hat{\bs{w}}_{k,\CA}}{\hat{u}^2_k} + \frac{\hat{\bs{w}}^H_{k,\CA}\!\left(\bs{H}_k  - \bs{H}^T_k\right) \hat{\bs{w}}_{k,\CA}}{\hat{u}_k}, \label{eq:OP2 c1 app}
\end{align}
which is obviously convex in $\bs{w}_{k,\CA}$ and $u_k$, where $\bs{H}_k = \bs{h}^H_{k,\CA}\bs{h}_{k,\CA}$. 

By using \eqref{eq:OP2 c1 app} as an approximation of \eqref{eq:OP dc1 c1}, problem \eqref{OP:dc1} can be approximated as
\begin{align}
\mathrm{P3}(\hat{\bs{w}},\hat{\bs{u}}):& ~\underset{\bs{w}, \bs{u}}{\mathtt{Maximize}}
~~ \bar{B}{\sum}_{k=1}^K \log_2( 1 + u_k)  \label{OP:2 app} \\
\mathtt{s.t.} 
&~~ \eqref{eq:OP dc1 c2};~ \eqref{eq:OP dc1 c3};~\eqref{eq:OP2 c1 app}. \notag
\end{align}

For given $\hat{\bs{w}}_{k,\CA}, \hat{x}_k$, the objective function in \eqref{OP:2 app} is concave and the constraints are convex, hence it can be solved in an efficient manner by standard solvers, e.g., CVX. Because the right-hand-side of \eqref{eq:OP2 c1 app} is always less than or equal to $\frac{{\bs{w}}^H_{k,\CA}\bs{H}_k {\bs{w}}_{k,\CA}}{{u}_k}$, the approximated problem \eqref{OP:2 app} always gives a suboptimal solution of the original problem \eqref{OP:dc1}.

In order to reduce the performance gap between the approximated problem \eqref{OP:2 app} and the original problem \eqref{OP:dc1}, we propose Algorithm~\ref{algo:dc} which consists of solving a sequence of SCA problems. The premise behind the proposed algorithm is to better select the parameters $\hat{\bs{w}}_{k,\CA}, \hat{u}_k$ through iterations.
\begin{algorithm}
	\caption{\textsc{Iterative Algorithm to solve \eqref{OP:dc1}}}\label{algo:dc}
	\begin{algorithmic}[1]
		\State Initialize $\hat{\bs{w}}_{k,\CA},\hat{u}_k$, $\epsilon$, $X_{\rm old}$ and $\mathtt{error}$.
		\State \textbf{while} $\mathtt{error} > \epsilon$ \textbf{do} 
		\State  \qquad Solve problem $\mathrm{P3}(\hat{\bs{w}}_{k,\CA},\hat{u}_k)$ in \eqref{OP:2 app} to obtain  $\bs{w}^\star_k, u^\star_k, \forall k$
		\State \qquad Compute $\mathtt{error} = |\bar{B}\sum_{k=1}^K\log_2(1 + u^\star_k) - X_{\rm old}|$
		\State \qquad Update $X_{\rm old} \gets \bar{B}\sum_{k=1}^K\log_2(1 + u^\star_k)$;~ $\hat{\bs{w}}_{k,\CA} \gets \bs{w}^\star_k$;~$\hat{u}_k \gets u^\star_k, \forall k$
	\end{algorithmic} 
\end{algorithm}

%

\begin{remark}[Initialization of Algorithm~\ref{algo:dc}]
Finding a feasible point is always essential in the SCA. Intuitively, one can think about the feasibility problem of \eqref{OP:1}, which is stated as
\begin{align}
&\underset{\{\bs{w}_{k,\CA}\}}{\mathtt{Maximize}}
~~ 1 \label{OP:P3 feasibility}\\
&\mathtt{s.t.} 
~~ \frac{1}{\bar{\eta}_k} |\bs{h}_{k,\CA} \bs{w}_{k,\CA}|^2 \ge \sum_{i \neq k} |\bs{h}_{k,\CA} \bs{w}_{i,\CA}|^2 + \sigma^2,\forall k, \subeqn \label{eq:p3 fea c1}\\
&\qquad~~{\sum}_{k=1}^K \|\bs{w}_{k,\CA} \|^2\leq P_{tot}. \subeqn \label{eq:p3 fea c2}
\end{align}
However, since both sides of \eqref{eq:p3 fea c1} are convex, this constraint is unbounded. Therefore, finding a feasible point by solving \eqref{OP:P3 feasibility} is not efficient. Instead, we adopt \eqref{OP:SDR1 ini} as the mean to find initial values $\hat{\bs{w}}, \hat{\bs{u}}$. In particular, from $\bs{W}_{\star k}, \forall k$, the solution of the convex problem \eqref{OP:SDR1 ini}, we obtain the corresponding feasible precoding vectors $\bs{w}_{\star k}$. Then, we assign  $\hat{\bs{w}}_k = \bs{w}_{\star k}$ and $\hat{u}_k = \frac{|\bs{h}_{k,\CA} \bs{w}_{\star k}|^2}{{\sum}_{i \neq k} |\bs{h}_{k,\CA} \bs{w}_{\star i}|^2 + \sigma^2}$.

\end{remark}

\subsection{JASPD Algorithm and Complexity Analysis}
Once the precoding vectors have been optimized for each antenna subset, i.e., problem \eqref{OP:1} is solved, we can tackle the original optimization problem \eqref{OP:0} via Algorithm~\ref{algo:JASPA}.
%
%

The proposed JASPD algorithm consists of two loops: the outer loop tries all valid antenna subsets, and the inner loop optimizes the precoding vectors iteratively. While the complexity of the inner loop is relatively reasonable since (the SDR of) problem \eqref{OP:SDR1 app} (or problem \eqref{OP:2 app}) is convex \cite{Boyd_OptimizationTime}, the outer iteration's complexity increases combinatorially with the number of antennas. In fact, the JASPD has to examine all $\binom{N}{M}$ candidates for the selected antennas. As an example, for $N = 20, M = 8$, there are $125970$ possible antenna subsets to be went through, each of which imposes an inner loop in Algorithm~\ref{algo:SDR} or Algorithm \ref{algo:dc}. 
%
Although guaranteeing the maximal achievable rate, the proposed JASPD suffers an exponential complexity due to the selection process. Its high computation time may limit its applicability in practice and degrade the effective rate (see \eqref{eq:R_k}). In the next section, we propose a low-complexity joint design to overcome the computation burden of the antenna selection process. 
\begin{algorithm}[t]
	\centering
	\caption{\textsc{Exhaustive Search based Joint Antenna Selection and Precoding Design }}\label{algo:JASPA}
	\textbf{Inputs}: $\bs{H}, P_{tot}, \{\eta_k\}_{k=1}^K$. \textbf{Outputs}: $C_{opt}, \CA_{opt}, \bs{W}_{opt}$	
	\begin{algorithmic}[1]
		\State Construct the super group $\CAb = \{\CA~|~ \CA \subset [N], |\CA| = M\}$
		\State Initialize $C_{opt} = 0$
		\State \textbf{for} $i=1:|\CAb|$ \textbf{do}
		\State \qquad	$\CA = \CAb[i]$ 
		\State \qquad	Apply Algorithm~\ref{algo:SDR} or Algorithm~\ref{algo:dc} on the current antenna subset $\CA$ 
		to obtain the optimal $X_{old}(\CA)$ and $\bs{W}_\star(\CA)$
		\State $\mathtt{If}~ C_{opt} < X_{old}(\CA)$ \\
		\qquad$C_{opt} \gets X_{old}(\CA)$; $\CA_{opt} \leftarrow \CA$; $\bs{W}_{opt} = \bs{W}_\star(\CA)$.
	\end{algorithmic} 
\end{algorithm}
%
%
%
%
%
\section{Accelerating the Optimization: A Deep Learning-based Approach}

In this section, we exploit recent advances in machine learning to overcome the major high-complexity limitation of selection process by proposing a learning-based antenna selection and precoding design algorithm (L-ASPD). 
%
%
%
The premise behind the proposed L-ASPD is to exploit machine-learning based predictions to help the optimal algorithm to tackle the most difficult and time-consuming part in the optimization. In particular, the L-ASPD will first predict potential subsets of antennas, which will be much smaller than $\binom{N}{M}$. 

\begin{figure}[]
	\centering
	\includegraphics[width = 0.6\columnwidth]{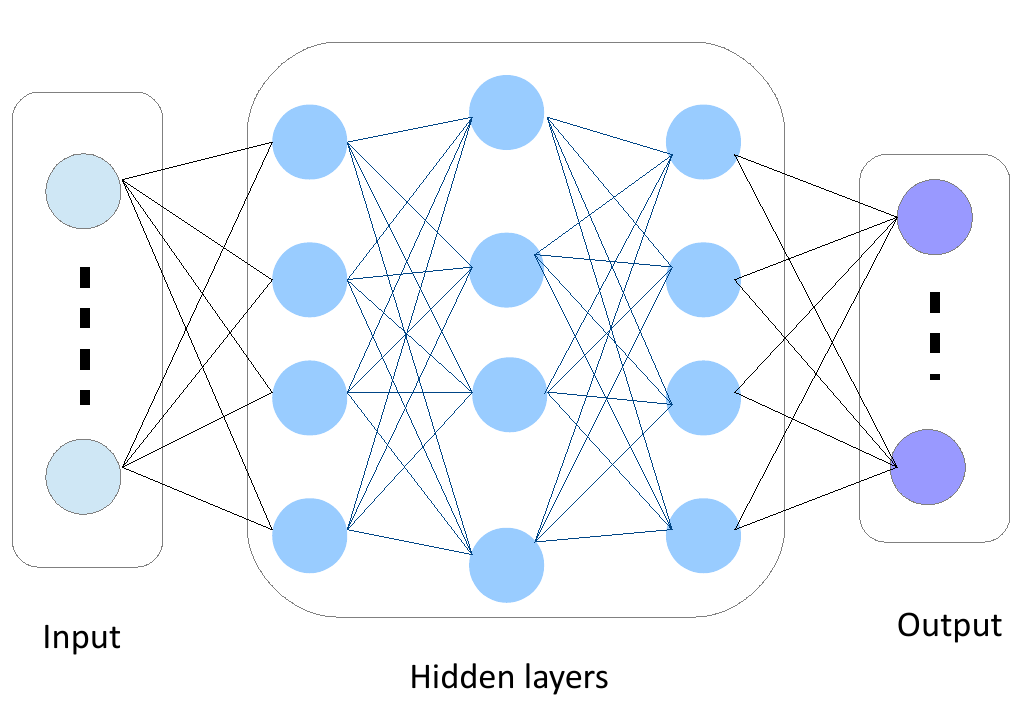}
	\caption{Illustration of a DNN with three hidden layers.}
	\label{fig:NN}
\end{figure}

We deploy DNN as the learning model to establish underlaying relations between the system parameters (inputs) and the selected antenna subset. The DNN consists of three main parts: one input layer, one output layer and hidden layers, as depicted in Fig.~\ref{fig:NN}. Based on the labeled data, the DNN will optimize the learning parameters in order to minimize the prediction error, e.g., cost function. 
The L-ASPD is implemented via 3 steps: i) offline training data generation, ii) building the learning model, and iii) real-time prediction.

\subsection{Training Data Generation}
Since the communication between the BS and the users is specified by the channel gains, the transmit power budget and noise power, they are essential for the learning model. Let $\bs{H} = [\bs{h}_1^H, \dots, \bs{h}_K^H]^H \in \mathbb{C}^{K\times N}$ denote the channel coefficients from the BS's antennas to all users. Since the number of users can be arbitrary between 1 and $M$ (the number of RF chains), the channel matrix  $\bs{H}$ is first zero-padded to obtain the standard size $\bar{\textbf{H}} = [\bs{H}^H, \boldsymbol{0}_{N\times (M-K)}]^H \in \mathbb{C}^{M \times N}$. Because the NN accepts only real-value inputs, the original complex representation of the channel matrix is invalid. One can stack the real and imaginary parts of $\bar{\textbf{H}}$ and use them as the training input to the NN \cite{AS_multicast_Ibrahim2018}. However, we observe that such method is not efficient to our problem because it does not directly capture inter-user interference - the major limiting factor in multiuser systems. As the inter-user interference is determined by the cross-product of the channel vectors of two users, we choose $\bs{x} = \frac{P_{tot}}{\sigma^2}\mathrm{abs}(\mathrm{vec}(\bar{\textbf{H}} \bar{\textbf{H}}^H)) \in \mathbb{R}^{M^2\times 1}$ as the training input. It is worth noting that the training input $\bs{x}$ is robust against the number of users and pathloss, as well as the BS's transmit power. Last but not least, $\bs{x}$ should be normalized before being fed to the NN, i.e., $\bs{x} = \frac{\bs{x}}{\max(\bs{x})}$.

Once the input sample is given, we need to define the output, which is the selected antenna combination that provides the maximum objective function in \eqref{OP:0}. For each training input $\bs{x}$, we define an output vector $\bs{b} \in \{0,1\}^{\binom{N}{M}\times 1}$ that consists of all possible antenna subsets. $b[n] = 1$ if the $n$-th subset is selected, otherwise $b[n] = 0$. Because we are interested in selecting only one subset, we have $\|\bs{b}\|_0 = 1$. In order to compute $\bs{b}$, for each channel realization $\bs{H}$ (corresponding to $\bs{x}$), we run the proposed JASPD algorithm to find the best antenna subset $\CA^\star$ and then assign the output element $b[n^\star] = 1$ corresponding to $\CA^\star$. 
\begin{table}[]
	\caption{\textsc{Steps to generate training samples for L-ASPD}}\label{table:training}
	\vspace{-0.2 cm}
	\begin{tabular}{l l}
		\hline		
		1.& \textbf{For} $t = 1:N_S$ \\
		2.& \qquad Generate a random number of users $K$ between $[1, M]$. \\
		3.& \qquad Generate random locations of these $K$ users between 50 and \\
		& \qquad 300m from the BS. Calculate the pathloss.\\
		4. & \qquad Generate a channel matrix $\bs{H}\in\mathbb{C}^{K\times N}$, including \\
		&\qquad the pathloss.\\	
		&\quad \textbf{Output sample generation}\\
		5.& \qquad Run JASPD algorithm to find the best antenna subset.	\\
		6.&\qquad Compute the binary output vector $\bs{b}_t$ with only a single \\
		&\qquad  non-zero element corresponding to the selected subset. \\
		&\quad \textbf{Input sample generation}\\
		5. & \qquad Zero-padding: $\bar{\textbf{H}} = [\bs{H}^H, \boldsymbol{0}_{N\times (M-K)}]^H$. \\
		6. & \qquad Calculate $\bs{x}_t = \frac{P_{tot}}{\sigma^2}\mathrm{abs}(\mathrm{vec}(\bar{\textbf{H}}^H \bar{\textbf{H}}))$; $\bs{x}_t = \frac{\bs{x}_t}{\max(\bs{x}_t)}$. \\
		7.& \textbf{Endfor}\\
		\hline
	\end{tabular} 
\end{table}
Denote by $N_S$ the number of samples used to train the learning model. The total training input is aggregated in the input matrix $\bs{X} = [\bs{x}_1, \bs{x}_2, \dots, \bs{x}_{N_S}]$, where $\bs{x}_t$ is the $t$-th input sample. Similarly, the training output matrix is $\bs{B} = [\bs{b}_1, \dots, \bs{b}_{N_S}]$, where $\bs{b}_t$ is the $t$-th output sample corresponding to the input sample $\bs{x}_t$. The steps for generating the training samples are listed in Table~\ref{table:training}. We note that JASPD algorithm considered in Table~\ref{table:training} is used for generating training samples and is executed off-line. Once the NN is well-trained, it is used for only the selected antenna subsets in the real-time prediction phase.

\subsection{Building the Learning Model}
When the training data is available, it will be used to train the NN with the learning parameter $\bs{\Theta}$. For an $L$-layer NN, we have $\bs{\Theta} = [\bs{\theta}_1, \dots, \bs{\theta}_L]$, where $\bs{\theta}_l \in \mathbb{R}^{N_l\times 1}, 1\le l \le L$, is the learning parameters in the $l$-th layer, and $N_l$ is the number of nodes in the $l$-th layer. As the most popular and efficient candidate for classification problems, we employ a sigmoid-family $\mathtt{tansig}(z) =  2(1+e^{-2z})^{-1}-1$ as the activation function for the hidden layers and the soft-max as the activation function for the output layer. The learning phase can be done via the minimization of prediction error
\begin{align}
\Delta(\bs{\Theta}) =& \frac{1}{N_S} \|-\mathrm{Tr}(\bs{B}^T \log(f_{\bs{\Theta}}(\bs{X})))  \\
-& \mathrm{Tr}(\bar{\bs{B}}^T \log(1 - f_{\bs{\Theta}}(\bs{X})))\parallel^2 
+ \frac{\lambda}{2N_S} {\sum}_{l=1}^L \parallel \bs{\theta}_l\parallel^2, \notag
\end{align}
where $\lambda$ is the regulation parameter, $\bar{\bs{B}} = \bs{1} - \bs{B}$, and $f_{\bs{\Theta}}(\bs{X})$ is the prediction of the output layer.

\subsection{Real-time Prediction}
When the NN has been well trained, it is ready to provide real-time and highly accurate predictions. From the current channel coefficient matrix $\bs{H}$, we construct $\bs{x} = \frac{P_{tot}}{\sigma^2}\mathrm{abs}(\mathrm{vec}(\bar{\textbf{H}}^H \bar{\textbf{H}}))$, where $\bar{\textbf{H}}= [\bs{H}^H, \boldsymbol{0}_{N\times (M-K)}]^H$, which is then normalized to obtain $\bs{x}_{\rm norm} = \frac{\bs{x}}{\max(\bs{x})}$. Then $\bs{x}_{\rm norm}$ is used as the input of the trained NN to output the prediction vector $\hat{\bs{b}}$. It is worth noting that the NN does not provide absolute prediction, e.g., $0$ or $1$, but probabilistic uncertainties, e.g., $-1\le \hat{b}[n] \le 1, \forall n$. In general, the larger an element in $\hat{\bs{b}}$ is, the higher chance this element is the best antenna subset. Consequently, the subset $\CA_{n^\star}$ corresponding to the largest output prediction, i.e., $n^\star = \arg\max_n \hat{b}[n]$, can be selected. 
\begin{algorithm}[t]
	\caption{Proposed L-ASPD Algorithm}\label{algo:4}
	\textbf{Inputs}: \bs{$\Theta$}, $\bs{H}, P_{tot}, \{\eta_k\}_{k=1}^K$. \textbf{Outputs}: $C_{opt}, \CA_{opt}, \bs{w}_{opt}$
	\begin{algorithmic}[1]
		\State Construct $\bs{x} = \frac{P_{tot}}{\sigma^2}\mathrm{abs}(\mathrm{vec}(\bs{H}^H \bs{H}))^2$; $\bs{x}_{\rm norm} = \frac{\bs{x}}{\max(\bs{x})}$
		\State  Apply $\bs{x}_{\rm norm}$ to the learned model $\bs{\Theta}$ to predict $\bs{\mathcal{K}}_S$
		\State Initialize $C_{opt} = 0$
		\State \textbf{for} $\CA \in \boldsymbol{\mathcal{K}}_S$ 
		\State ~~ Apply Algorithm ~\ref{algo:SDR} or \ref{algo:dc} on the current subset $\CA$ to \\
		~~ obtain the optimal $X_{old}(\CA)$ and $\bs{w}_{\star,\CA}$
		\State ~~ $\mathbf{if}~ C_{opt} < X_{old}(\CA)$ 
		\State ~~~~~~ $C_{opt} = X_{old}(\CA)$; $\CA_{opt} \leftarrow \CA$; $\bs{w}_{opt} \gets \bs{w}_{\star ,\CA}$.
	\end{algorithmic}
\end{algorithm}
However, the prediction is not always precise. Therefore, in order to improve the performance of L-ASPD, instead of choosing only one best candidate, we select $K_S$ subsets, denoted by $\boldsymbol{\mathcal{K}}_S$, corresponding to the $K_S$ largest elements in $\hat{\bs{b}}$. Then, we apply the precoding design (Algorithm~\ref{algo:SDR} or \ref{algo:dc}) on these $K_S$ subsets. Intuitively, larger values of $K_S$ will increase the chance for the L-ASPD to select the best antenna subset at an expense of more computation complexity. The steps of the L-ASPD are listed in Algorithm~\ref{algo:4}. Compared with the JASPD, the L-ASPD significantly reduces the computational time since it tries only $K_S$ promising candidates instead of $\binom{N}{M}$. Consequently, the L-ASPD is expected to achieve higher effective sum rate than that of the JASPD, especially when $K_S \ll \binom{N}{M}$.

%


\section{Performance Evaluation}
In this section, we evaluate the performance of the proposed algorithms via simulation results. The users are uniformly distributed in an area between $50$ and $300$ meters from the centered-BS. We employ the WINNER II line-of-sight pathloss model \cite{WinnerII}, which results in that the pathloss is uniformly distributed between $-59.4$ dB and $-74.6$ dB. All wireless channels are subject to Rayleigh fading. The channel bandwidth $B = 1$ MHz and the noise spectral density is -140 dBm/Hz. 
We adopt the LTE specifications \cite{LTEwhitepaper} that one c.u. lasts in one symbol duration and is equal to 66.7 $\mu$s, and one block duration is spanned over 200 c.u.. The BS is assumed to spend 0.2 c.u. to solve one convex optimization problem \cite{Boyd_OptimizationTime}. As a result, it takes $0.2K_S$ c.u. to execute the proposed L-ASPD, where $K_S$ is the number of predicted subsets. We employ an NN with two hidden layers to train the learning model for the L-ASPD, each layer consists of 100 nodes\footnote{We heuristically try a different number of hidden layers and find out that a NN with two hidden layers is sufficient for our problem}. SVM can also be employed for its fast training phase, however, results in poorer performance compared to NN. This is because SVM results in hyperplanes to discriminate the data whereas the NN can discriminate data using more elaborate functions. The NN is trained using the scaled conjugate gradient method. Other simulation parameters are listed in Table~\ref{tab:para}.

\begin{table}[!h]
	\begin{center}
	\caption{Simulation parameters}\label{tab:para}
	\begin{tabular}{|l||l|} 
		\hline 
		\textbf{Parameters} & \textbf{Value} \\[0.5ex]
		\hline 
		Cell radius & 300 m\\
		BS's transmit power & 1 W - 5W \\
		Number of RF chains $M$ & 4 \\
		Number of antennas $N$ & Varies\\
		Number of users $K$ & Varies between 1  and $M$\\
		QoS $\eta_k = \eta, \forall k$ & 2 Mbps \\
		Training method & Scaled conjugate gradient\\
		Activation function (hidden layers) & $\mathtt{tansig}$\\
		Activation function (output layer) & $\mathtt{soft-max}$\\
		Loss function & Cross-entropy\\[1ex] 		
		\hline
	\end{tabular}
\end{center}
\end{table}

\begin{figure}[]
	\centering
	\subfigure[Sum rate versus number of iterations]{\includegraphics[width = 0.48\columnwidth]{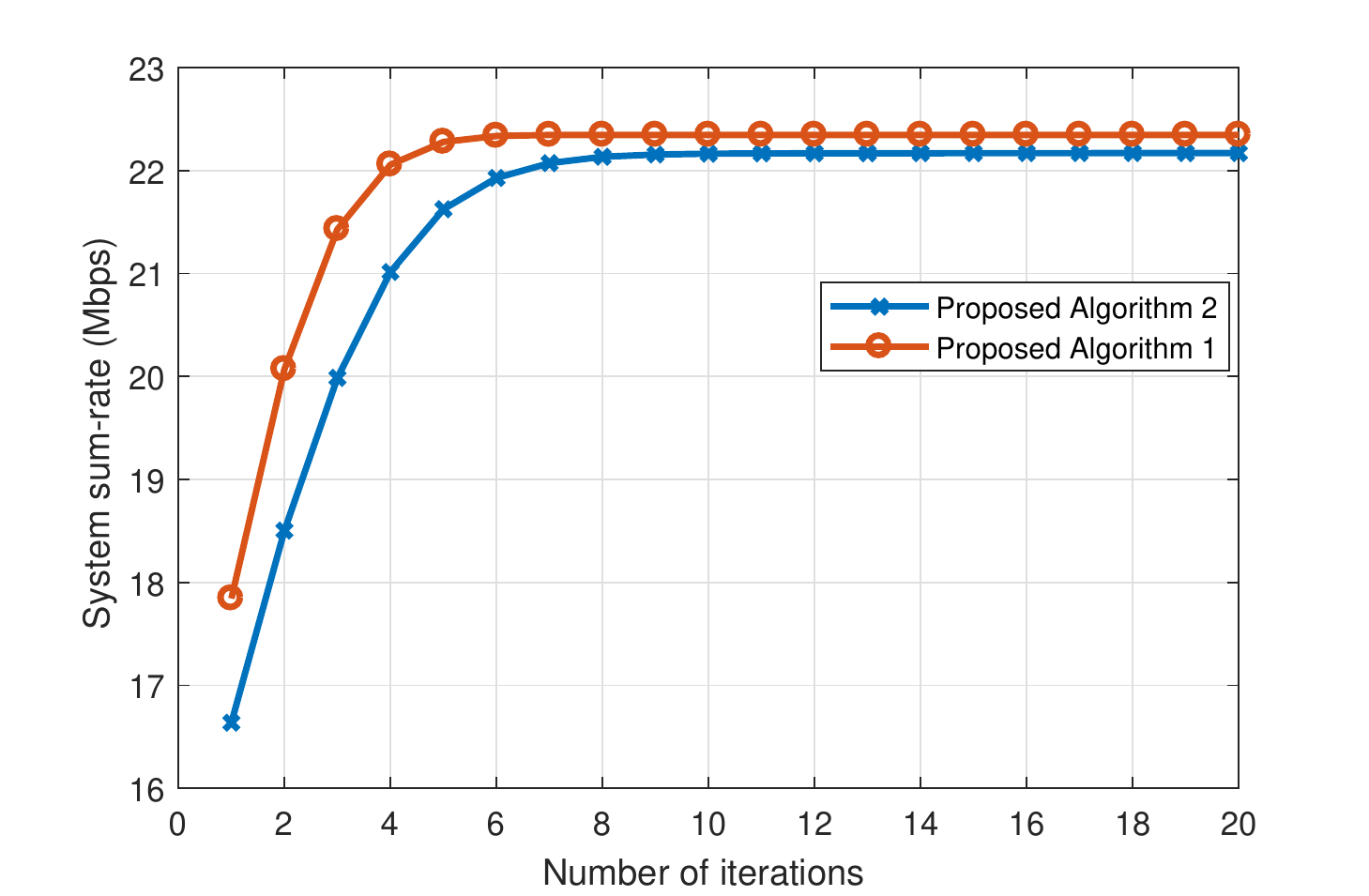}}
	\subfigure[Sum rate versus simulation time]{\includegraphics[width = 0.48\columnwidth]{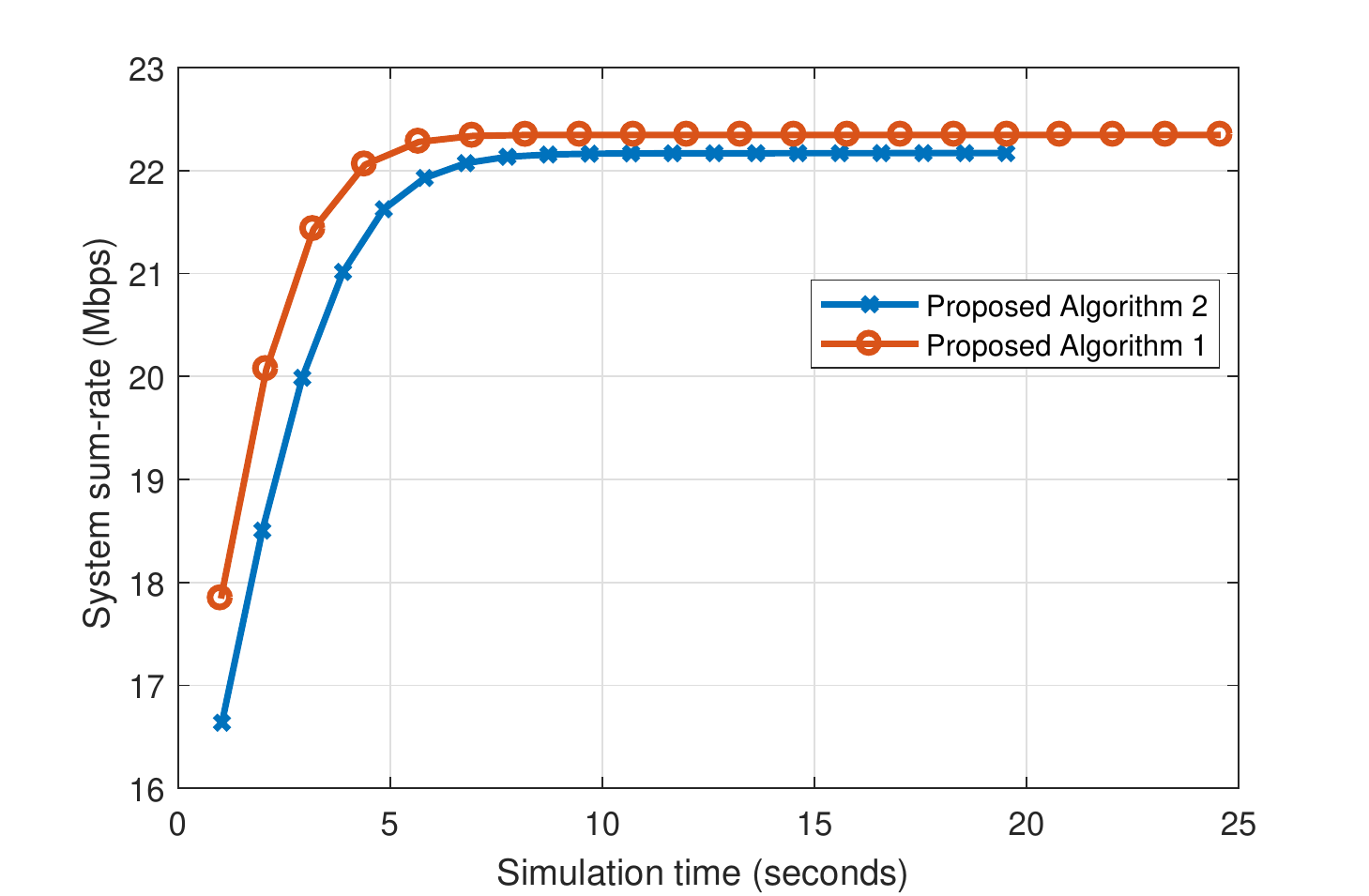}}
	\caption{Performance comparison of the proposed Algorithm 1 and 2, $P_{tot} = 37$ dBm and $K = 4$. Both algorithms converge in less than 10 iterations.}
	\label{fig:convergence}
\end{figure}

\subsection{Convergence of the Proposed Optimization Algorithms}
We first evaluate the convergence performance of the proposed iterative Algorithm \ref{algo:SDR} and \ref{algo:dc} presented in Section~\ref{sec:JASPD}. The results are obtained from 200 random realizations of channel fading coefficients and users' locations. For each realization, we run both Algorithm~\ref{algo:SDR} and \ref{algo:dc} until they converge. Fig.~\ref{fig:convergence}a compares the sum-rate obtained by the two proposed algorithms as a function of the iteration number. It is clearly shown that both algorithms converge quickly after less than 10 iterations, which demonstrates the effectiveness of the proposed iterative algorithms. 

In order to provide insights on the computation performance of the proposed algorithms, we show in Fig~\ref{fig:convergence}b the sum-rate versus the simulation time. Both algorithms are carried out by SeDuMi solver integrated in Matlab 2017b, running on a personal laptop with the Intel i7-6820HQ CPU and 8GB RAM. It is observed that Algorithm~\ref{algo:dc} executes slightly faster than Algorithm~\ref{algo:SDR}, however, achieves a smaller sum-rate. The performance gain brought by Algorithm~\ref{algo:SDR} results from the fact that it uses more memory than Algorithm~\ref{algo:dc}, as shown in Table~\ref{tab:1}. Due to superior performance, we will employ the proposed Algorithm~\ref{algo:SDR} in the remaining comparisons.
\begin{table}[!h]
	\centering
	\caption{\textsc{Number of variables required by Algorithm 1 and 2 for different setups for $N = 8$.}}\label{tab:1}
	\begin{tabular}{|l||l|l|l|l|} 
		\hline 
		\textbf{$M$} & $2$ & $3$ & $4$ & $5$ \\
		\hline 	
		Algorithm 1 & 267 & 400 & 533 & 666 \\
		Algorithm 2 & 55 & 94 & 141 & 196 \\
		\hline
	\end{tabular}
\end{table}

\subsection{Performance-complexity Trade-off of the L-ASPD}
In this subsection, we examine the efficiency of the proposed L-ASPD via a performance-complexity gain trade-off. By confining the search space of the prediction output, i.e., $K_\CS$ - the number of potential antenna subsets, we can manage the complexity of L-ASPD since it will work only on $K_\CS$ candidates. The complexity gain of L-ASPD is defined as the relative time saving compared to the exhaustive search that tries every antenna subsets, calculated as:
\begin{align}
	\theta(K_\CS) = \frac{\tau(\binom{N}{M} - K_\CS)}{\tau \binom{N}{M}} = 1 - \frac{K_\CS}{\binom{N}{M}},
\end{align} 
where $\tau$ is the computational time spent on the optimization of the precoding vectors for a selected antenna subset. The performance gain is defined as the ratio between the sum rate obtained by L-ASPD divided by the optimal sum rate which is achieved by searching all possible antenna subsets.

\begin{figure}[]
	\centering
	\includegraphics[width = 0.6\columnwidth]{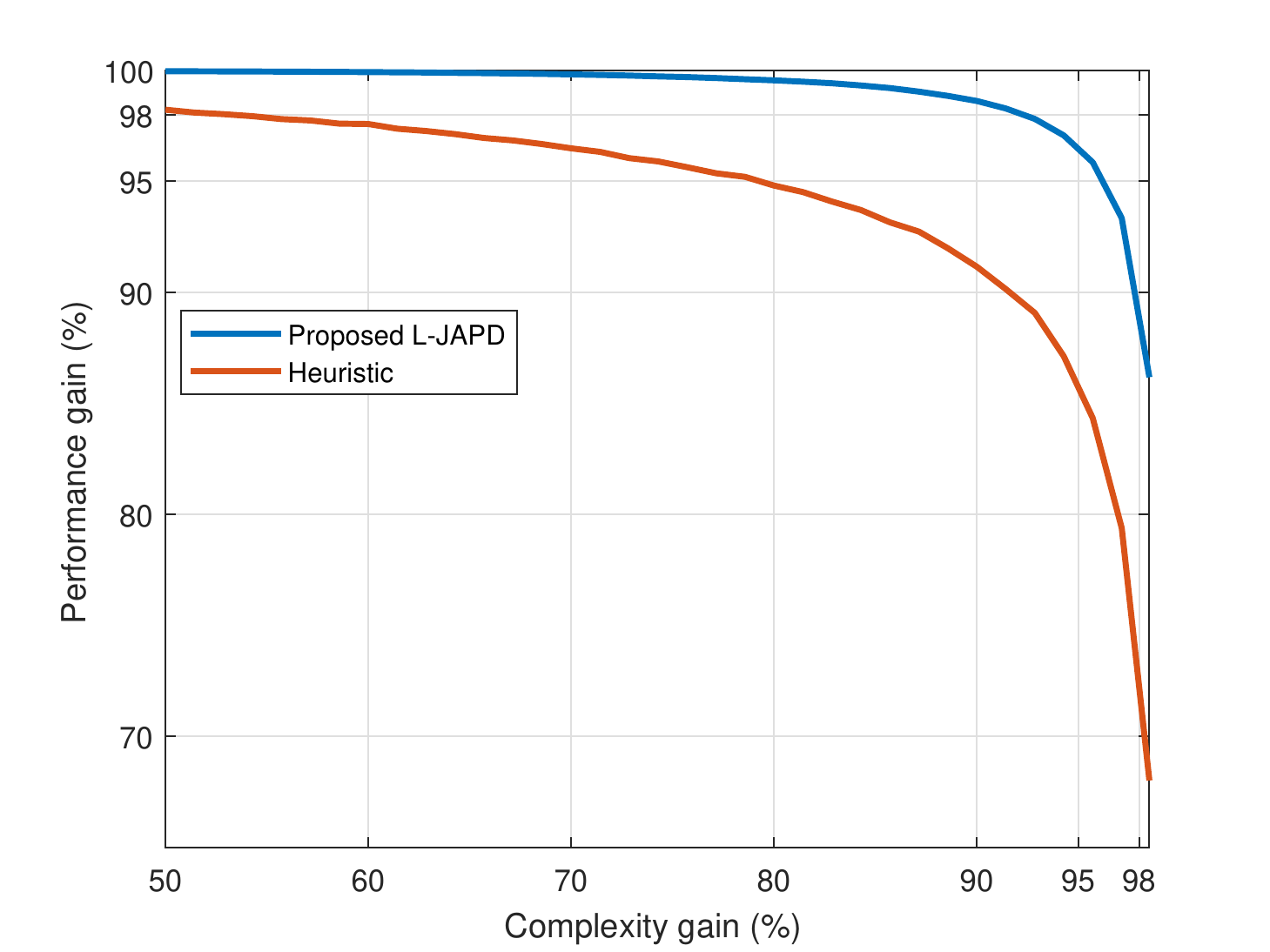}
	\caption{Performance-complexity tradeoff of the proposed L-ASPD. $ M = 4, N = 8$.}
	\label{fig:tradeoff}
\end{figure}

Fig.~\ref{fig:tradeoff} plots the performance-complexity tradeoff of the proposed L-ASPD with $M=4$ RF chains and $N=8$ total number of antennas. It is observed that the L-ASPD retains more than 96\% of the optimal sum rate (which is obtained by exhaustive search) while saving more than 95\% complexity. Even when spending only 2\% the computational time, the L-ASPD still achieves 86\% the optimal performance, which confirms the effectiveness of the proposed L-ASPD algorithm. Compared with the heuristic solution, the L-ASPD further reduces more than 13\% the computational time at the 95\% performance gain target. 

\begin{figure}
  	\centering
  	\includegraphics[width=0.6\columnwidth]{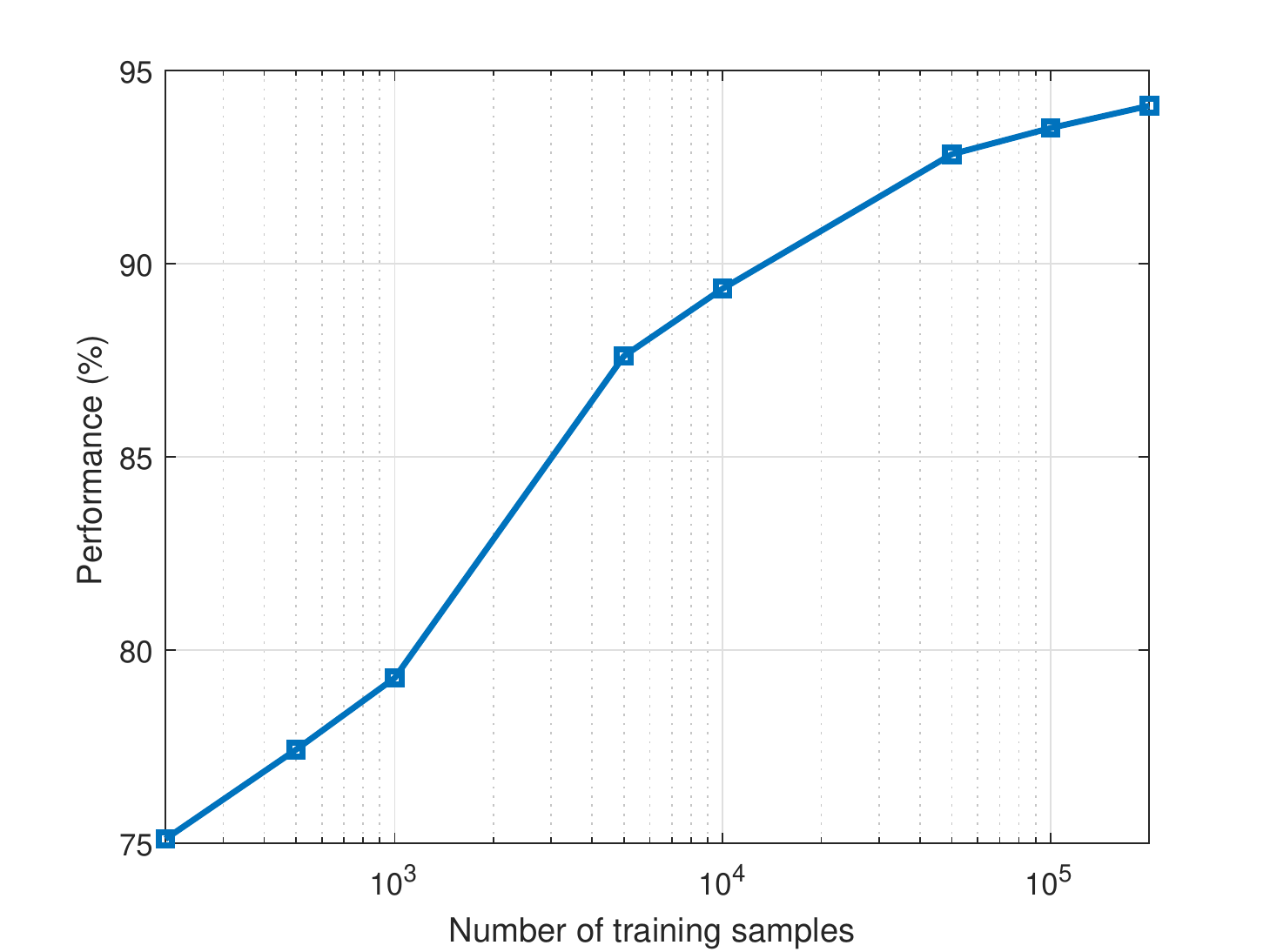}
  	\caption{Learning (relative) performance versus the number of training samples. $M = 4, N = 8$.}
  	\label{fig:vsSample}
  	\vspace{-0.2cm}
  \end{figure}

Fig.~\ref{fig:vsSample} plots the relative performance in the real-time prediction of L-ASPD versus the number of training samples. The relative performance is measured as the ratio of the L-ASPD's sum rate divided by the one obtained by the JASPD. Each training sample is generated randomly and captures the randomness in both channel small-scale fading and user location. In general, having more training samples results in better prediction accuracy since the L-ASPD learns more about the intrinsic relation between the selected antennas and the input features. It is shown that $2\times 10^5$ training samples are sufficient for the L-ASPD to achieve more than 94\% of the optimal performance. 


\subsection{Online Performance Comparison}

This subsection demonstrates the effectiveness of the proposed L-ASPD algorithm via performance comparisons with existing solutions in difference scenarios. The first baseline scheme is proposed in \cite{Chen07_AntenSelection}, which employs block diagonalization to consecutively eliminate antennas that incur the largest transmit power cost. The second baseline is introduced in  \cite{AS_multicast_Ibrahim2018}, which is a learning-assisted antenna selection for multicasting. In addition, a \emph{Heuristic} search is also presented, which also applies the proposed beamforming design but it searches for the antenna subset heuristically. We note that comparison with \cite{ML20_AS_HybridBeamformer,jingong,AS_wiretap_He2018} is not applicable  because \cite{ML20_AS_HybridBeamformer,jingong} consider a single-user system and \cite{AS_wiretap_He2018} selects only a single antenna.

\begin{figure}
	\centering
	\includegraphics[width=0.6\columnwidth]{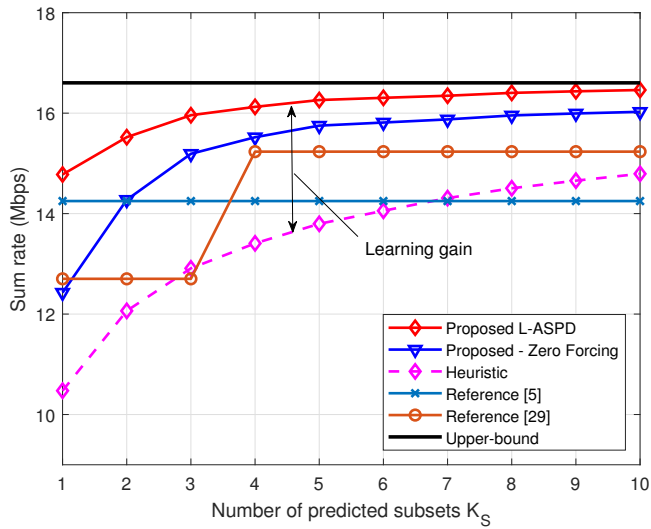}
	\caption{Sum rate performance of the proposed algorithms versus the number of predicted subsets $K_S$. $P_{tot} = 33$ dBm, $M = 4$ and $N = 8$.}
	\label{fig:online}
\end{figure}
Fig.~\ref{fig:online} shows the achievable sum rate as a function of $K_S$ - the most promising subsets predicted by the proposed L-ASPD algorithm. In order to reveal the benefit of proposed beamforming design in Algorithm~\ref{algo:SDR}, we also show a curve, which applies a zero-forcing based power control \cite{VuIcc18} on the antenna subsets predicted by  Algorithm~\ref{algo:4}. This curve is named as \emph{Proposed - Zero Forcing} in the figures. It is shown that the proposed L-ASPD significantly surpasses all schemes for all observed $K_S$ values. In general, having more predicted subsets $K_S$ results in a larger sum rate, which is in line with results in Fig.~\ref{fig:tradeoff}. In particular, by searching over the most five promising subsets, the proposed L-ASPD achieves 1 Mbps and 2 Mbps higher than schemes in \cite{AS_multicast_Ibrahim2018} and \cite{Chen07_AntenSelection}, respectively. We note that the sum rate of the scheme in \cite{Chen07_AntenSelection} is independent from $K_S$ since it predicts the best antenna subset. Similarly, the performance curve of  \cite{AS_multicast_Ibrahim2018} has a step-shape because it uses the active antennas as the prediction outputs, hence it is only able to confine the original search space to $\binom{M+n}{M}$ subsets, with $0\le n\le N-M$.  

\begin{figure}
	\centering
	\includegraphics[width=0.6\columnwidth]{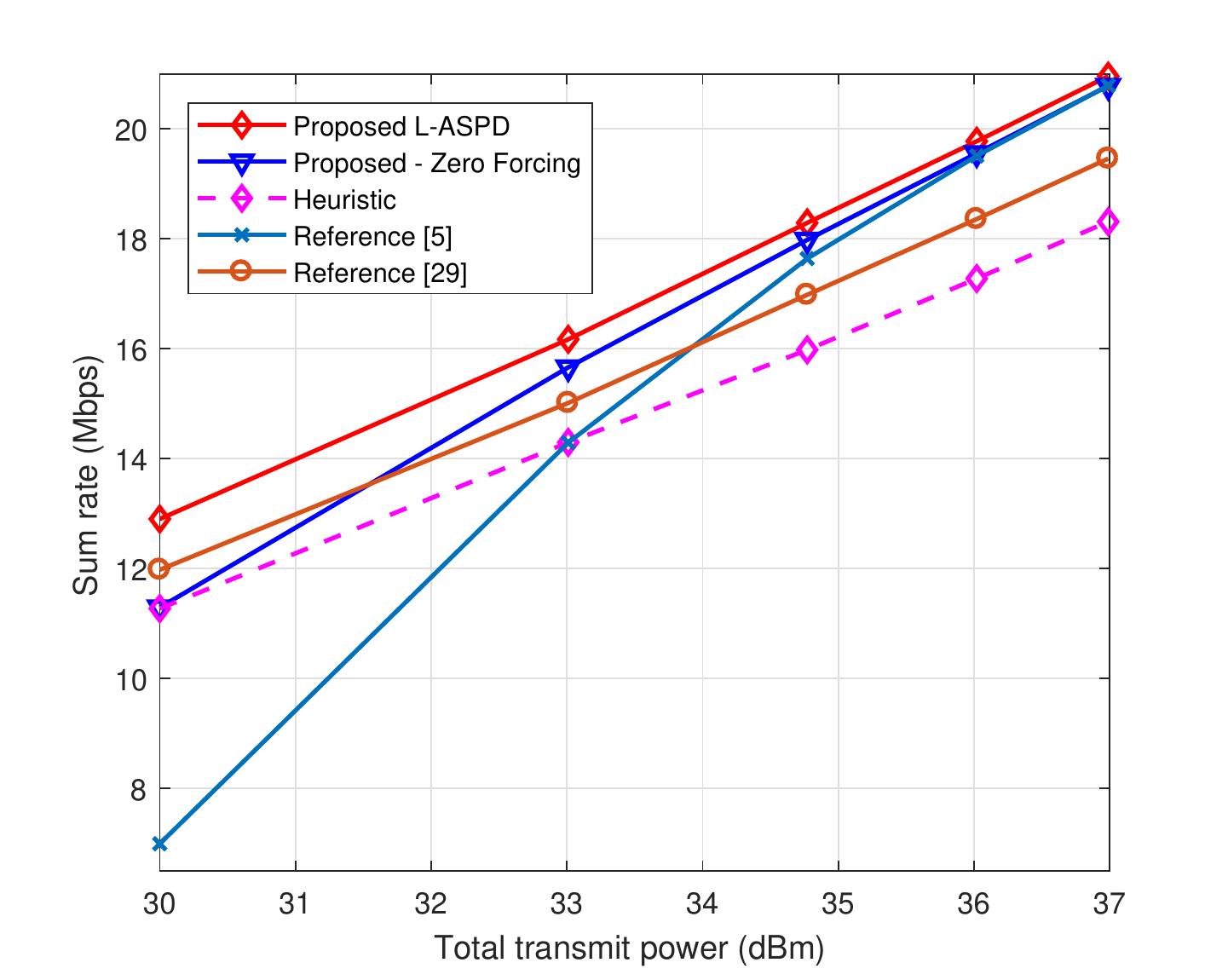}
	\caption{Sume rate performance of the proposed algorithms versus the total transmit power $P_{tot}$. $K_S = 7$ and $N = 8$ available antennas.}
	\label{fig:vsPtot}
\end{figure}

Fig. \ref{fig:vsPtot} plots the sum rate as a function of the transmit power. The effectiveness of the proposed learning-based method is shown via the largest sum rate achieved by the L-JAPD compared to other schemes. On average, the L-JAPD algorithm produces 1.5 Mbps and 2 Mbps more than the solution in  \cite{AS_multicast_Ibrahim2018} and heuristic scheme, respectively, proving that the NN has been well trained. Compared to the solution in \cite{Chen07_AntenSelection}, the L-ASPD achieves a relative sum rate gain of 5 Mbps and 2 Mbps at the transmit power equal to 30 dBm and 33 dBm, respectively. One interesting observation is that the Zero-forcing scheme and the solution in \cite{Chen07_AntenSelection} approach the L-ASPD's performance when the total transmit power budget increases. This is because for large $P_{tot}$, the BS has sufficient power budget to fully mitigate inter-user interference. For small $P_{tot}$, the system resource becomes scarce, therefore completely eliminating inter-user interference is far from the optimum, which is shown in a big gap between the L-ASPD and these two schemes. In such high-load scenarios, employing the proposed design is highly beneficial.

\begin{figure}
	\centering
	\includegraphics[width=0.6\columnwidth]{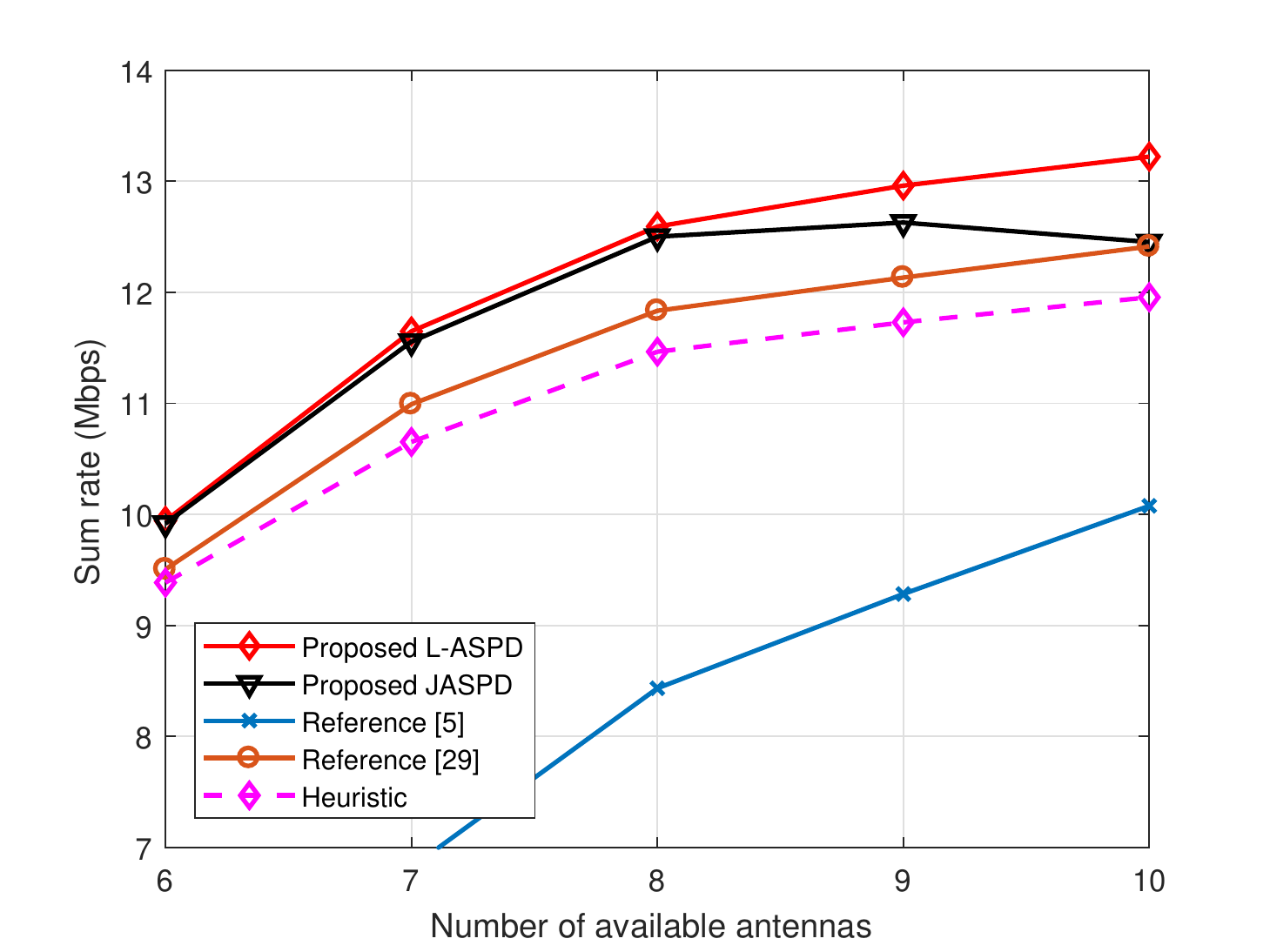}
	\caption{Effective sum rate comparison for various number of total antennas $N$. $P_{tot}$ = 30 dBm, $M = 4, K_S = 10$.}
	\label{fig:vsN}
\end{figure}
Fig.~\ref{fig:vsN} presents the effective sum rate for different total antennas numbers $N$. For a fair comparison, the total transmit power is kept constant at 30 dBm and the total overhead of channel estimation and computation is taken into account. For the former, it takes 8 c.u. to obtain the CSI when the total antenna number is $6, 7, 8$, and takes 12 c.u. when the number of antennas is 9 and 10. Consider the latter, the L-ASPD algorithm only searches over 10 most promising candidates, while the JASPD tries all $\binom{N}{M}$ antenna subsets. In general, having more antennas results in higher effective sum rate of all schemes, which confirms the benefit of antenna selection. Interestingly, the proposed L-ASPD algorithm achieves the best performance and surpasses the exhaustive search scheme, especially for large $N$, which is in contrast to common understanding that the exhaustive search achieves the best performance. This is because we take the computation time into account in the comparison, as shown in \eqref{eq:R_k}. As a result, the exhaustive search scheme spends too much time in searching for the best subset, particularly with large $N$, resulting in smaller effective rates. As an example for $N = 10$, the exhaustive search scheme requires a computation time which is 21 times more than that of the L-ASPD.

\section{Conclusions}
We studied the joint design for antenna selection and precoding vectors in multi-user multi-antenna systems to fully exploit the spatial diversity. We first proposed a (near) optimal joint antenna selection and precoding algorithm to maximize the system sum rate, subjected to the users' QoS and limited transmit power. The proposed joint design successively optimizes the precoding vectors via two proposed iterative optimization algorithms based on the semidefinite relaxation and successive convex approximation methods. In order to further improve the optimization efficiency, we then developed the machine learning-based solution to provide appropriate and time-stringent antenna predictions. The proposed learning-based algorithm is robust against the number of users and their locations, the BS's transmit power, as well as the channel fading. We showed via simulation results that the proposed learning-based solution significantly surpasses existing selection schemes and outperforms the exhaustive search-based solution. 

Based on the outcome of this work, several research directions can be considered. The first problem is how to improve the training phase efficiency, which is especially important when the number of available antennas is very large. In such a case, a low-complexity precoding design, e.g., zero-forcing, can be used to quickly obtain sufficient training samples. The second problem lies in dealing with the network dynamics, which requires the learning model to frequently and timely adapted. Transfer leaning and reinforcement learning are promising solutions in this case to avoid retraining the whole network.



\appendices
\section{Proof of Proposition \ref{prop SDR}} \label{app:prop SDR}
Denote $\big(\bs{W}^{(t)}_\star, \bs{x}^{(t)}_{\star}, \bs{y}^{(t)}_{\star}\big)$ as the optimal solution of $\text{P2}(\bs{y}^{(t)}_0)$ at iteration $t$. We will show that if $y_{\star k}^{(t)} < y^{(t)}_{0k}, \forall k$, then by using $y^{(t+1)}_{0k} = y_{\star k}^{(t)}$ in the $(t+1)$-th iteration, we will have $\sum_k x_{\star k}^{(t+1)}  > \sum_k x^{(t)}_{\star k} $, where $\{x_{\star k}^{(t+1)}\}_{k=1}^K$ is the solution at iteration $t+1$. Indeed, by choosing a relatively large initial value $\bs{y}^{(1)}_0$, we always have $y_{\star k}^{(1)} < y^{(1)}_{0k}, \forall k$. 

Denote $f(y;a) = e^a(y - a + 1)$ as the first order approximation of the $e^y$ function at $a$. At iteration $t+1$, we have $y^{(t+1)}_{0k} = y^{(t)}_{\star k}, \forall k$. Therefore, $f(y;y_{\star k}^{(t)})$ is used in the right-hand side of constraint \eqref{eq:OP SDR1 app c1} at the $(t+1)$-th iteration. Consider a candidate $(y^{(t+1)}_1, \dots, y^{(t+1)}_K)$ for any $y^{(t+1)}_k \in (\hat{y}_k, y^{(t)}_{\star k})$, where $\hat{y}_k = y_{\star k}^{(t)} - 1 + e^{y^{(t)}_{0k} - y_{\star k}^{(t)}}(y_{\star k}^{(t)} - y^{(t)}_{0k} + 1)$. Because function $\exp()$ is convex and $y^{(t+1)}_k <  y_{\star k}^{(t)}$, then we have $f(y^{(t+1)}_k; y_{\star k}^{(t)}) > f(y_{\star k}^{(t)}; y^{(t)}_{0k}), \forall k$. Therefore, there exits $\bs{W}^{(t+1)}_k$ and $x^{(t+1)}_k > x^{(t)}_{\star k}$ which satisfies constraints \eqref{eq:app1} and \eqref{eq:OP SDR1 app c1}. Consider a new set $\{\bs{W}^{(t+1)}_k, x^{(t+1)}_k, y^{(t+1)}_k\}_{k=1}^K$. This set satisfies all the constraints of problem $\text{P2}(\bs{y}^{(t)}_\star)$, and therefore is a feasible solution of the optimization problem. As the result, the optimal objective at iteration $(t+1)$, $\frac{\bar{B}}{\log(2)}\sum_k x^{(t+1)}_{\star k}$, must satisfy $\frac{\bar{B}}{\log(2)}\sum_k x^{(t+1)}_{\star k} \ge \frac{\bar{B}}{\log(2)}\sum_k x^{(t+1)}_k > \frac{\bar{B}}{\log(2)}\sum_k x^{(t)}_{\star k}$, which completes the proof of Proposition 1.

\section{Convexity of function $\frac{\bs{x}^T\bs{A}\bs{x}}{y}$}\label{app 1}
To prove the convexity of $F(\bs{x},y) = \frac{\bs{x}^T\bs{A}\bs{x}}{y}$ for any positive semi-definite matrix $\bs{A}$, we need to show that the Hessian matrix of $F(\bs{x},y)$ is positive semidefinite. Indeed, the Hessian matrix of $F(\bs{x},y)$ is 
\begin{small}
	\begin{align}
	\bs{H}_F = \left[
	\begin{array}{c c}
	\frac{\bs{A} + \bs{A}^T}{y} & -\frac{(\bs{A} + \bs{A}^T)\bs{x}}{y^2}\\
	-\frac{\bs{x}^T(\bs{A} + \bs{A}^T)}{y^2} & \frac{2\bs{x}^T\bs{A}\bs{x}}{y^3} 
	\end{array}
	\right]. \notag
	\end{align}
\end{small} 

For arbitrary vector $\bs{c} = [\bs{a}^T b]^T$, where $\bs{a} \in \mathbb{R}^{N\times 1}$, consider a function
\begin{small}
	\begin{align}
	&\bs{c}^T \bs{H}_F \bs{c} = \frac{\bs{a}^T(\bs{A} + \bs{A}^T)\bs{a}}{y} - \frac{\bs{a}^T(\bs{A} + \bs{A}^T)\bs{x}b}{y^2} \notag \\
	&\qquad\qquad
	- \frac{\bs{x}^T(\bs{A} + \bs{A}^T)\bs{a}b}{y^2} + \frac{2\bs{x}^T\bs{A}\bs{x}b^2}{y^3} \notag \\
	&\overset{(*)}{=}\frac{\bs{a}^T\!(\bs{A}\! +\! \bs{A}^T)\bs{a}}{y}\! -\! 2\frac{\bs{a}^T(\bs{A}\! +\! \bs{A}^T)\bs{x}b}{y^2} 
	+ \frac{\bs{x}^T(\bs{A}\! +\! \bs{A}^T)\bs{x}b^2}{y^3} \notag\\
	&= \frac{\bs{a}^T\tilde{\bs{A}}\bs{a} - 2\bs{a}^T\tilde{\bs{A}}\tilde{\bs{x}} + {\tilde{\bs{x}}}^T\tilde{\bs{A}}\tilde{\bs{x}}}{y},\label{eq:HF}
	\end{align}
\end{small}
where $\tilde{\bs{A}} \triangleq \bs{A}^T + \bs{A}$, $\tilde{\bs{x}} \triangleq  \bs{x}b/y$ and $(*)$ results from the fact that $\bs{A}$ is symmetric and $\bs{a}^T\tilde{\bs{A}}\tilde{\bs{x}} = \tilde{\bs{x}}^T\tilde{\bs{A}}\bs{a}$. It is obvious that the RHS of \eqref{eq:HF} is always non-negative for $y > 0$ and positive semi-definite matrix $\tilde{\bs{A}}$, which concludes the positive semi-definite of the Hessian matrix of $F(\bs{x},y)$.

\end{document}